\documentclass{article}

    \PassOptionsToPackage{numbers, compress}{natbib}

    \usepackage[preprint]{neurips_2023}

\usepackage[utf8]{inputenc} 
\usepackage[T1]{fontenc}    
\usepackage{hyperref}       
\usepackage{url}            
\usepackage{booktabs}       
\usepackage{amsfonts}       
\usepackage{nicefrac}       
\usepackage{microtype}      
\usepackage{bm}
\usepackage{braket}
\usepackage[dvipdfmx]{graphicx}
\usepackage[dvipdfmx]{color}

\title{LCAONet: Message-passing with physically optimized atomic basis functions}

\author{
  Kento Nishio \\
  School of Engineering \\
  The University of Tokyo \\
  Tokyo, 113-8656 \\
  \texttt{knishio@iis.u-tokyo.ac.jp} \\
  \And
  Kiyou Shibata, Teruyasu Mizoguchi\thanks{http://www.edge.iis.u-tokyo.ac.jp} \\
  Institute of Industrial Science \\
  The University of Tokyo \\
  Tokyo, 153-8505 \\
  \texttt{\{kiyou,teru\}@iis.u-tokyo.ac.jp} \\
}

\begin{document}

\maketitle

\begin{abstract}
A Model capable of handling various elemental species and substances is essential for discovering new materials in the vast phase and compound space.
Message-passing neural networks (MPNNs) are promising as such models, in which various vector operations model the atomic interaction with its neighbors.
However, conventional MPNNs tend to overlook the importance of physicochemical information for each node atom, relying solely on the geometric features of the material graph.
We propose the new three-body MPNN architecture with a message-passing layer that utilizes optimized basis functions based on the electronic structure of the node elemental species.
This enables conveying the message that includes physical information and better represents the interaction for each elemental species.
Inspired by the LCAO (linear combination of atomic orbitals) method, a classical method for calculating the orbital interactions, the linear combination of atomic basis constructed based on the wave function of hydrogen-like atoms is used for the present message-passing.
Our model achieved higher prediction accuracy with smaller parameters than the state-of-the-art models on the challenging crystalline dataset containing many elemental species, including heavy elements.
Ablation studies have also shown that the proposed method is effective in improving accuracy.
Our implementation is available online.
\end{abstract}

\section{Introduction}

Machine learning (ML) has recently attracted attention in computational materials as an alternative to the high computational cost of first-principles methods such as density functional theory (DFT). 
By using ML method, we can accurately simulate large systems for which first-principles calculations are not applicable, and thus gain insight that is closer to real materials that have vacancies, substitution sites, and so on.
Many ML methods have been developed using features (descriptors) that encode geometrical, topological, physical, and chemical information about the local environment of the atoms.
The most important aspect of descriptor construction is its uniqueness in representing a material \cite{Faber2015-ba, MBTR2017}.
In addition to the topology, physical and chemical information such as orbital-orbital interactions, atomic charges, and electric dipoles are also crucial for the functional manifestation of a material.
Therefore, developing descriptors that adequately represent these properties is essential for accurate modeling, but it is not easy \cite{Ghiringhelli2015-xg}.
The development and selection of descriptors require a lot of manual coordination and years of experience and background knowledge about the material science.

Against this background, a number of deep learning methods based on Message-Passing neural networks (MPNNs), which can extract the local environment of atoms without requiring manual adjustments through message-passing architecture, have been developed and shown promising results in predicting quantum mechanical properties of molecular and crystalline systems.
In the message-passing layer, geometric information of a material, such as interatomic distances and bond angles, are converted into vectors with physically motivated basis functions, and the graph embedding vectors are updated by vector operations.
In most cases, the functional form of the basis functions for vectorizing the geometry is the same regardless of the node elemental species.
However, in actual materials, even if the structure is the same, the atomic interaction should be different depending on the type of neighboring atoms. In addition, such message-passing methods can capture geometric information but not physical and chemical properties of elemental species. Therefore, extracting physically rich representations such as descriptors through message-passing layers is currently difficult.

In this work, we propose an improved method of three-body MPNN that optimize the form and the number of basis functions according to the electronic structure of node atoms.
Based on a new set of basis functions developed by the physical formulation of atomic interactions, we develop a physically motivated weighted message-passing that can propagate interactions between orbitals inspired by the LCAO-MO method, a classical computational method for modeling orbital interactions (LCAONet).
We also develop a method to generate chemically richer initial node vectors by embedding the number of electrons in the ground-state occupied orbitals in addition to the atomic number.
The proposed improved methods are investigated by ablation studies, and it is shown that each method contributes significantly to the accuracy improvements.
Modeling that incorporates physicochemically rich representations such as descriptors without human intervention via a message-passing architecture outperforms existing ML methods.
It is suggested that in order to develop a universal model that can handle a variety of elemental species, it may be useful to include more inductive bias in the model according to node atoms.

\section{Related work}

\paragraph{Descriptor based models.}
ML methods for predicting material properties have traditionally used descriptors, which contain geometrical, physical and chemical information.
SOAP \cite{SOAP2013, SOAP2016}, for example, encodes the local environment by expanding the atomic density based on the spherical harmonics and radial basis functions.
OFM \cite{Lam_Pham2017-dz} encodes the interaction between valence orbitals of the neighboring atoms based on the electron configuration.
A concept common to many of these descriptors is the encoding of information about the system based on "nearsightedness" \cite{Prodan2005-dq}.
After atomic numbers and atomic positions of the system is transformed using manually adjusted descriptors, it is input into the Gaussian process \cite{Bartok2010-oe, CoulombMatrix2012, Bartok2017-vi} or neural networks \cite{Behler2007, Behler2011, Smith2017-rt, Gastegger2018-pz, HDNNP2021-uo}.
Although descriptors have a sophisticated expression, their hyperparameters need to be well tuned to the system to develop an accurate model.

\paragraph{MPNNs for molecules.}
Graph neural networks (GNNs) were first proposed in the 90s \cite{Baskin1997-jr, Sperduti1997-fy} and 00s \cite{Gori2005-ul, Scarselli2005-jw, Scarselli2009-fc} and then applied to molecular graphs as message-passing neural networks \cite{NIPS2015_Divenaud, Gilmer2017-mu}.
When the GNN is applied to material graphs, the geometric features of the graph are vectorized by physically motivated basis functions, which are used to model interactions with neighboring atoms.
The geometric features such as the distance between two nodes \cite{CGCNN, SchNet,  MEGNet, PhysNet}, the angle between two edges consisting of three nodes \cite{DimeNet_2020, DimePP_2020, M3GNet_2022-up}, and also the dihedral angle generated between four nodes \cite{GemNet_2021, GemNetOC_2022} are used.
In this study, from the viewpoint of computational cost, message-passing was performed including the three-body interaction.
Message-passing with rotational equivariant features has also been developed to efficiently incorporate higher-order geometric features \cite{tensorfield2018, Cormorant2019, painn2021, sphericalMPNN2021}.
Although these studies have demonstrated excellent accuracy for molecular and crystalline dataset, the basis function form is constant regardless of the node elemental species.
In this study, we provide a new physically motivated message-passing framework to select basis functions based on the electronic structure of the node atom.

\paragraph{MPNNs with physicochemical information.}
There have been efforts to incorporate physical information such as orbital interactions into MPNNs.
In OGCNN \cite{OGCNN2020-og}, OFM is incorporated during the generation of graph embeddings, and message-passing incorporates orbital interactions into the node embedding vectors.
However, OGCNN only deals with valence orbital information, and this study incorporates richer orbital information, including inner-shell orbitals, and higher-order geometric features up to three-body interactions.
In OrbNet \cite{OrbNet}, orbital features are generated in advance by the semi-empirical DFT calculation, and the graph-attention model is constructed using pre-computed features.
In this study, since the goal is to learn the topology and physical interactions from the graph only, no additional computed numerical features were added to the model.
However, it is possible to add such precomputed features to this model.
In SpookyNet \cite{spookynet2021}, the spin state of the molecular system and all electrons are added as inputs and long-range interactions are taken into account in the message-passing layer to achieve high accuracy for molecular systems.
In this study, following SpooKyNet, the electronic structure information of each atom is used as the embedding.
Furthermore, the difference in messages depending on the spin state is included in the basis function.

\section{LCAO-MO method}
The Linear Combination of Atomic Orbital - Molecular Orbital (LCAO-MO) method is a calculation approach that approximates the molecular orbital or, more broadly, the electronic state of a system as a linear combination of atomic orbitals that are strongly bounded for each atom.
Due to its simplicity, the LCAO-MO method has long been used in the analysis of many molecular systems \cite{mcquarrie1997physical}.

Here, we apply the LCAO-MO method to the simplest system, hydrogen molecular ion ($\mathrm{H_2^+}$), and outline the methodology.
Hydrogen molecular ion is the simplest system consisting of two atomic nucleus with one nucleon number and one electron.
Consider approximating the molecular orbitals (MO) of the entire system by the LCAO-MO method.
Note that in hydrogen molecular ion, the Schr\"{o}dinger equation can be solved annalytically under the Born-Oppenheimer approximation \cite{Born} but here we obtain an approximate solution to understand the LCAO-MO method.

Let $\phi$ be the 1s orbital of the one-electron wave function bounded around each nucleus, then the MO $\psi$ of $\mathrm{H_2^+}$ approximated by the LCAO-MO is as follows:

\begin{equation}
    \psi = c_1 \phi_{A} + c_2 \phi_{B}.
\end{equation}

Each coefficient can be obtained from the variational principle; Determining the coefficients so that the energy $E$ obtained from $\psi$ and the Schr\"{o}dinger equation $H\psi=E\psi$ is minimized.
Energy is expressed using MO as follows:

\begin{equation}
    E=\frac{\Braket{\psi|H|\psi}}{\Braket{\psi|\psi}}=\frac{\sum_i \sum_j c_i^{*} c_j H_{ij}}{\sum_i \sum_j c_i^{*} c_j S_{ij}},
\end{equation}

where $S_{ij}=\Braket{\psi_i|\psi_j}$, $H_{ij}=\Braket{\psi_i|H|\psi_j}$, and $c_i$ is determined from the minimum energy condition $\frac{\partial E}{\partial c_i}=0$.
Determine the coefficients by solving the equations $(\bm{H}-E\bm{S})\bm{c}=0$ that is derived from minimization conditions.
Then the coefficients are determined, a relationship, $c_2/c_1=\pm1$, is derived between the two coefficients.
This means that the contribution of each nucleus is equal, which is a physically correct depiction.
The specific values of the coefficients can be obtained from the MO normalization conditions.

Thus, the LCAO-MO method is a very simple yet powerful method for gaining physical insight.

\section{Optimized basis functions with node elemental species}\label{basis}

\begin{figure}[bt]
    \centering
    \includegraphics[keepaspectratio, scale=0.13]{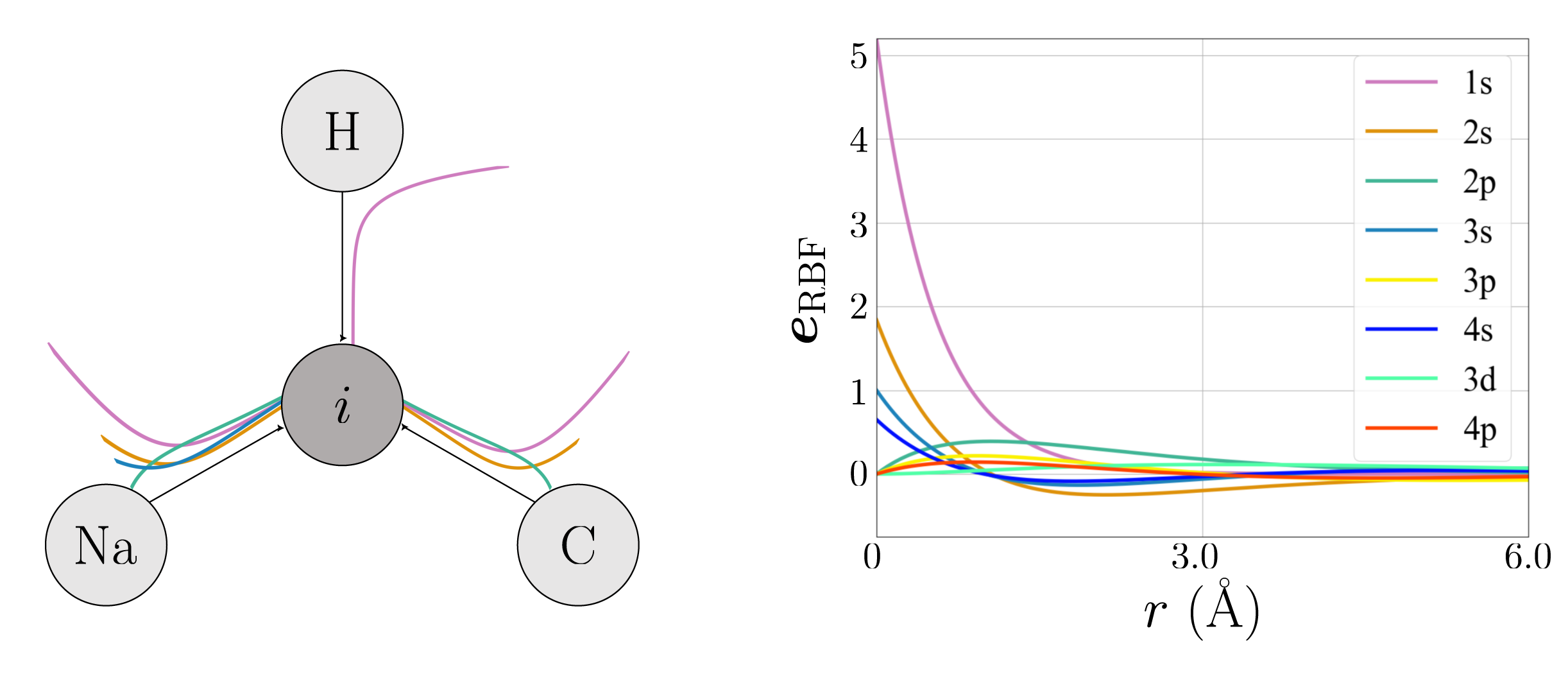}
    \caption{The message-passing scheme of this study, and the radial basis functions developed from the Schr\"{o}dinger equation for particles moving in the potential of the hydrogen-like atom.
    The radial basis function has more nodes as the value of $l$ increases (more nodes in $p$ and $d$ orbitals than in $s$ orbitals).
    The figure on the left shows how the number and type of basis functions are restricted according to elemental species.
    When considering message-passing for the central atom $i$, the hydrogen atom uses only the $1s$ orbital for message-passing. On the other hand, the Na atom uses the four basis functions $1s, 2s, 2p$ and $3s$.
    They all correspond to occupied orbitals in the ground state.
    }
    \label{fig:message-passing}
\end{figure}

\paragraph{Graph neural networks for materials.}
Material graphs can be uniquely represented by only pairs of atomic numbers $\bm{Z}=\{ z_1, \ldots, z_i, \ldots, z_N \}$ and atomic coordinates $\bm{X}=\{ \bm{x}_1,\ldots,\bm{x}_i,\ldots,\bm{x}_N \}$. $N$ represents the number of atoms constituting the system, and in the case of a periodic system, the number of atoms in the supercell is considered. GNNs input atomic numbers and atomic coordinates to predict quantum mechanical properties such as formation energy, highest occupied molecular orbital (HOMO) energy, dipole moment and so on. In this study, we predict the scalar regression target $t\in \mathbb{R}$, i.e., our GNN model is defined using the parameters $\theta$ via $f_{\theta}: \{ \bm{Z}, \bm{X} \} \rightarrow t$ as mentioned in DimeNet \cite{DimeNet_2020}.

\paragraph{Geometric representation.}
To represent the interaction between nodes, we use the pairwise interatomic distance $d_{ij}=||\bm{x}_i-\bm{x}_j||_2$, and the three-body angle $\alpha_{(kj,ji)}=\angle\bm{x}_k\bm{x}_j\bm{x}_i$ as geometric features. Then, a geometric feature vector $\bm{e}_{RBF}^{(ij)}, \bm{e}_{SBF}^{(ijk)}$ is created using basis functions and used for message-passing. As the radial basis function for vectorizing $\bm{e}_{RBF}^{(ij)}$, previous studies have used a set of Gaussian type radial basis functions \cite{CGCNN, SchNet, MEGNet, PhysNet}, but a set of spherical Bessel functions has been developed as an orthogonal basis more suitable for quantum systems and incorporated into many models \cite{DimeNet_2020, M3GNet_2022-up, painn2021}. In this study, we propose a radial basis function that can more accurately represent physical interactions based on the same spirit as the orthogonal basis proposed in DimeNet \cite{DimeNet_2020}. Furthermore, by varying the number and type of functions used depending on the electronic structure of the node elemental species, we improve parameter efficiency and bring useful inductive bias to the model. As the angular basis function for $\bm{e}_{SBF}^{(ijk)}$, normalized spherical harmonic function are used to vectorize as in the previous studies \cite{DimeNet_2020}. This follows naturally from the physical formulation that expresses the elemental interaction.

\paragraph{Physically motivated basis function.}
To construct the basis function based on the physical formulation, we consider a one-electron wave function bounded by a spherically symmetric potential, as in DimeNet.
In view of the fact that the model is to be constructed to reproduce the DFT results that calculate the electronic state using the electron density $\rho(\bm{r})=||\Psi(\bm{r})||^2$, it is desirable to construct the basis function based on the electron wave function $\Psi(\bm{r})$.
The electron wave function is defined by the time-independent Schr\"{o}dinger equation $\left( -\frac{\hbar^2}{2m}\nabla^2+V(\bm{r}) \right) \Psi(\bm{r}) = E\Psi(\bm{r})$, with constant mass $m$ and energy $E$.
Since it is difficult to know in advance the potential $V(\bm{r})$,  DimeNet derived the electron wave function by assuming a spherically symmetric potential that is 0 inside the cutoff sphere and $\infty$ outside the cutoff sphere.
The spherical Bessel function derived by the variable separation method was then used as the radial basis function, and the spherical harmonic function was used as the angular basis function.
However, the actual electrons are thought to be bound by the potential from the nucleus.
In other words, it is more natural to assume that the Coulomb potential $V(\bm{r})=-\frac{z_i e^2}{4\pi\epsilon_0}\frac{1}{||\bm{r}||}$ is working against the electrons in the cutoff sphere.
The form of the one-electron wave function in motion under the Coulomb potential of a nucleus is well known and can be obtained by the variable separation method in a polar coordinates $(r, \theta, \phi)$ \cite{Griffiths2018-uh}.

\begin{equation}
    \Psi_{nlm}(r, \theta, \phi) = R_{nl}(r)Y_m^l(\theta, \phi),
\end{equation}

where $n$, $l$, and $m$ are the principal, azimuthal, and magnetic quantum number, respectively.
Each quantum number is an integer satisfying the condition, $n=1,2,3 \dots$, $l=0,1,2\dots, n-1$ and $||m||\leq l$,
that determine the shape and orientation of the orbit.
For example, $n=1$, $l=0$ corresponds to the 1s orbital, while $n=2$, $l=1$ corresponds to the $2p$ orbital. 
$Y_m^l$ is the normalized spherical harmonic function, a spherically symmetric function, and $R_{nl}$ is a radial wave function with

\begin{equation}
    R_{nl}(r) = C_{nl} \left( \frac{2z_i}{na} \right) \left( \frac{2z_ir}{na} \right)^l \mathrm{exp}\left( -\frac{z_ir}{na} \right) L_{n+l}^{2l+1}\left( \frac{2z_ir}{na} \right),
\end{equation}

where $L_{n+l}^{2l+1}$ is associated Laguerre polynomials and $C_{nl}$ is the normalization factor, both functions of $n$ and $l$.
In this study, the $C_{nl}$ were determined by normalizing within the cutoff radius for crystalline systems and by normalizing over the whole space for molecular systems.
$r$ represents the distance from the center of the orbit i.e. atomic nucleus.
As described above, the spherical harmonic and the radial wave functions are derived from the  Schr\"{o}dinger equation of a particle moving around an hydrogen-like atomic potential, and are used as basis functions for vectorizing the graph.
In the next section, we will discuss how to optimize the basis functions for each node atom.

\paragraph{Optimization of basis functions according to elemental species.}
The number and type of basis functions are then optimized for each elemental species by using as the basis function only those radial wave functions corresponding to the orbitals occupied by the ground state, i.e., the orbitals in which electrons are present in the ground state (see Figure \ref{fig:message-passing}).
For example, for a carbon atom (C), the ground state electron configuration is  $(1s)^2(2s)^2(2p)^2$.
Message-passing is performed using only the basis corresponding to $1s, 2s, 2p$ and no other basis (e.g., $n = 3$ or more).
Note that such optimization restricts the basis function of, for instance, a hydrogen atom to only the $1s$ orbital, so that only messages that decay from the neighborhood are propagated.
In actual molecules and crystals, the spread of interactions is expected to be different for each elemental species, so the optimization of basis functions can more clearly represent the message differences among elemental species.
Since information on occupied orbitals is explicitly included in the basis, it is expected that information on the element's electronic structure and inter-orbital interactions can be included in the message passing.

\section{LCAONet}

\begin{figure}[bt]
    \centering
    \includegraphics[keepaspectratio, scale=0.32]{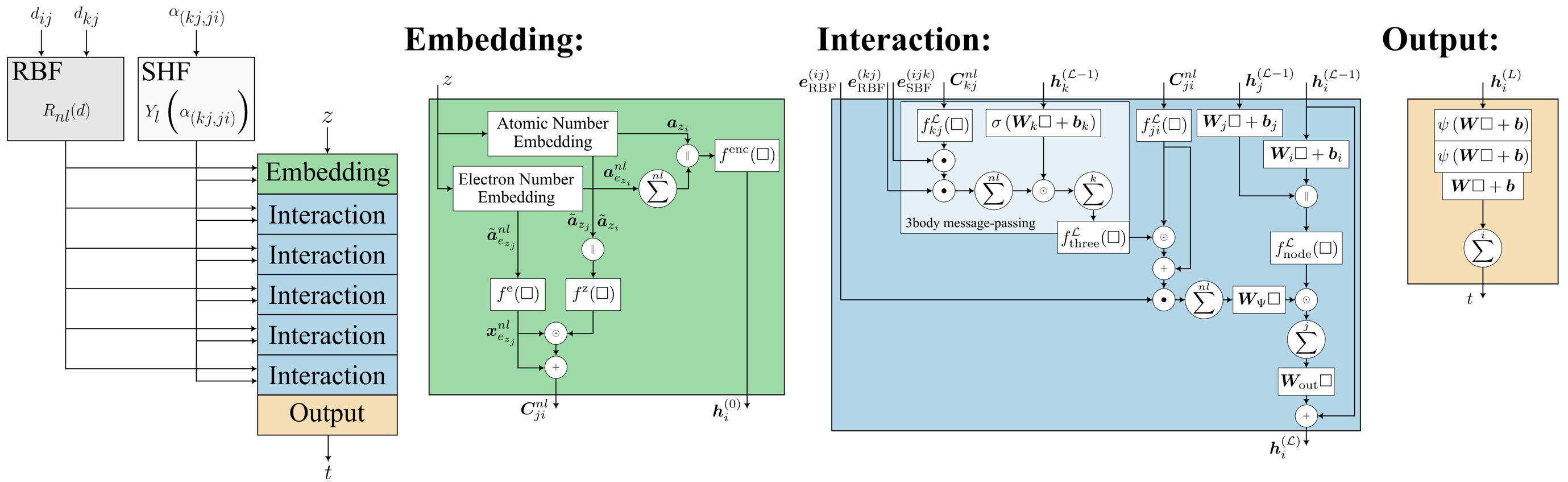}
  \caption{
    The LCAONet architecture. 
    $\Box$ denotes the layer’s input, $\parallel$ denotes concatenation, $\cdot$ denotes vector-scalar multiplication and $\odot$ denotes element-wise multiplication.
    The edge distance $d_{ij}$ and $d_{kj}$ is vectorized by the radial wave function of the hydrogen atom, and the three-body angle $\alpha_{(kj, ji)}$ is vectorized by the spherical harmonic function.
    In the embedding block, an initial embedding vector $\bm{h}^{(0)}_i$ is generated based on the atomic number and electronic structure information of the node atom $i$.
    In multiple interaction blocks, the message is propagated by the neighboring embeddings $\bm{h}_j$, the basis $\bm{e}_{\mathrm{RBF}}^{(ij)}$ optimized for each elemental species, and the basis coefficient vector $\bm{C}_{ji}^{nl}$, which updates the node embedding $\bm{h}_i$.
    In the output block, the embedding vectors are transformed to calculate node-wise scalar values, and finally they are summed up to output the predictions.
    }
  \label{fig:model}
\end{figure}

\subsection{Embedding block}\label{embed}

\paragraph{Electron embedding in the ground state.}
In this study, the number of electrons in the ground state is embedded in addition to the atomic number.
The number of electrons in each orbital is represented by a learnable, randomly initialized electron number weight vectors $\bm{a}^{nl}_{e_z}\in\mathbb{R}^F$ shared across the materials.
$n, l$ is the quantum number of the orbital and $e^{nl}$ is the number of electrons in the orbital of quantum number $n, l$ in the ground state, meaning that a different embedding is generated for each orbital. (For example, the embedding corresponding to the case where the 1s orbital has one occupied electron number is different from the embedding corresponding to the case where the 2s orbital has one occupied electron number.)

\paragraph{Embedding node and coefficient vectors.}
Atomic embedding $\bm{h}_i$ are initialized for each node, and the coefficient vectors $\bm{C}_{ji}^{nl}$ for adding the atomic bases are initialized for each edge.
These embedding vectors are initialized using the atomic number $z$ and the number of electrons of each orbital $e^{nl}$ as described above.
Atomic numbers are similarly embedded with a weight vector $\bm{a}_{z}\in\mathbb{R}^{F}$, which is learnable and randomly initialized.
This weight vector is also shared among materials.
We generate the initial node embedding vector $\bm{h}^{(0)}_i\in\mathbb{R}^{F_n}$ by combining orbital electron weights $\bm{a}_{e_z}^{nl}$ with atomic number weights $\bm{a}_z$ via

\begin{equation}
    \bm{h}^{(0)}_i = f^{\mathrm{enc}} \left( \bm{a}_{z_i} \parallel \sum^{nl}{\bm{a}^{nl}_{e_{z_i}}} \right),
\end{equation}

where $\parallel$ is the concatenation and $f^{\mathrm{enc}}$ is the encoding function.
In this study, a two-layer neural network was used to enable learning.

Similarly, the coefficient vector $C^{nl}_{ji}\in \mathbb{R}^{F}$ is also initialized using orbital electron weights $\bm{\tilde{a}}_{e_z}^{nl}\in\mathbb{R}^{F}$ and atomic number weights $\bm{\tilde{a}}_z\in \mathbb R^{F}$.
Here, the tilde on the atomic number and electron number weights indicates that a different parameter is used than the weights used for the node embedding.
In this study, a linear combination of each orbit is used for messages from the neighborhood. $C^{nl}_{ji}$ is used as the coefficient in the linear combination, but we want the coefficient vector of the orbit corresponding to the $n,l$ of the unoccupied orbit to be a zero vector.
To satisfy this conditions, initialize the coefficient vectors in three steps as follows.
First, the electron number weights $\bm{\tilde{a}}_{e_z}^{nl}$ are initialized with a zero vector when the number of the electron $e$ is zero (unoccupied orbital).
Even when learning parameters by back-propagation, the atomic basis used can be restricted, depending on the elemental species, as always a zero vector by eliminating gradient propagation.
Second, the weight of the orbital electron number is transformed by a certain function $f^{\mathrm{e}}$ to make $\bm{x}_{e_z}^{nl} \in \mathbb{R}^{F}$ i.e.
$\bm{x}_{e_z}^{nl} = f^{\mathrm{e}}(\bm{\tilde{a}}_{e_z}^{nl})$. In this process, the embedding $\bm{x}_{e_z}^{nl}$ is generated so that the vector of non-occupied orbitals becomes a zero vector.
To achieve this, $f^{\mathrm{e}}$ is a two-layer neural network without a bias term.
Finally, using an electron number embedding $\bm{x}_{e_z}^{nl}$ where only the vectors corresponding to the $(n, l)$ of the occupied orbitals have values, the coefficient vectors are initialized via

\begin{equation}
    \bm{C}^{nl}_{ji} = 
    \left\{  
        \bm{x}_{e_{z_j}}^{nl} + \bm{x}_{e_{z_j}}^{nl} \odot f^{\mathrm{z}}\left( \bm{\tilde{a}}_{z_i} \parallel \bm{\tilde{a}}_{z_j} \right)
    \right\},
\end{equation}

where $f^{\mathrm{z}}$ is a function that converts concatenated atomic number weights with learnable parameters.
$\odot$ represents the element-wise multiplication of the vectors, and the atomic number embedding is multiplied by the electron embedding $\bm{x}_{e_{z_j}}^{nl}$ to keep the unoccupied orbitals of neighboring $j$ atom in the zero vector.
$e_{z_j}$ represents the number of electrons in the neighboring atoms.
The reason for using the number of electrons in the $j$ atom to embed is that the information on the electronic structure and orbitals of the neighboring atoms is aggregated in the central atom $i$.
That is, $\bm{C}^{nl}_{ji}$ is modeled as how the $nl$ orbitals of the neighboring atom $j$ interact with the central atom $i$.

\subsection{Interaction block}

\paragraph{Message with electronic structure and orbital information.}
As described in \ref{basis}, the hydrogen-like wave functions, developed to more accurately model the atomic interactions, are used as atomic basis functions to vectorize the geometric features of the graph.
Then, inspired by the LCAO method, these linear combinations are used for message-passing, which includes electronic structure and orbital information.
Furthermore, by initializing the coefficient vectors of the linear combination as described in \ref{embed}, only the basis corresponding to the occupied orbitals is used for message-passing, and the number and type of basis functions are optimized according to the neighboring node elemental species.
The goal is thereby to capture inductive bias in the message according to atomic species.

\paragraph{Three body orbital interaction.}
First, directional message-passing\cite{DimeNet_2020} is performed based on the angle information consisting of the three-body nodes.
As in previous studies \cite{DimeNet_2020, DimePP_2020}, the geometric information is vectorized by a spherical harmonic basis with $m = 0$.
First, we create a linear combination of orbitals between the three-body to create the orbital message of $\mathcal{L}$ th layer $\Psi_{ijk}^{\mathcal{L}}\in \mathbb{R}^C$ via

\begin{equation}
    \Psi_{ijk}^{(\mathcal{L})}
    = \sum_{nl}R_{nl}(d_{kj}) Y_l^{0} \left( \alpha_{(kj, ji)} \right) \cdot f_{kj}^{\mathcal{L}}\left( \bm{C}^{nl}_{kj} \right),
\end{equation}

where $\cdot$ denotes vector-scalar multiplication and $f_{kj}^{\mathcal{L}}$ is a two-layer neural network, and as in \ref{embed}, the bias term is eliminated for the embedding of unoccupied orbitals is kept to $0$ vector.
Note that only the coefficient vectors of the occupied orbitals have non-zero values, so the basis used depends on the electronic structure of the $k$ atom.
For example, if the $k$ atom has only 1s orbitals, then only the basis corresponding to $n=1,l=0$ is used for the message.
This allows the message to include an appropriate spatial spread for each atom.
Since the three-body directional message-passing expresses the relative positional relationship between edges, in this study, this message aggregated with a coefficient vector that expresses the interaction of the edges:

\begin{equation}
    \tilde{\bm{C}}_{ji}^{nl} = 
    \bm{C}_{ji}^{nl} + 
    \bm{C}_{ji}^{nl} \odot 
    f_{\mathrm{three}}^{\mathcal L}
    \left (
        \sum_{k \in \mathcal{N}_j \setminus i}
            \sigma
            \left(
                \bm{W}_{k} \bm{h}^{(\mathcal{L})}_k + \bm{b}_{k}
            \right)
        \odot
        \Psi_{ijk}^{(\mathcal{L})}
    \right).
\end{equation}

Again, an element-wise product is taken with the three-body message and coefficient vectors and a residual network is used so that the unoccupied orbital coefficient vector is kept to 0 vector.

\paragraph{Edge orbital interaction.}
For nodes between two-body, the linear combination of the radial basis functions $\Psi_{ji}^{(\mathcal{L})} = \sum_{nl}R_{nl}(d_{ij}) \cdot f_{ji}^{\mathcal{L}}\left( \bm{C}^{nl}_{ji} \right) \in \mathbb{R}^C$ is also used for message-passing.
We update node embedding with orbital messages containing electronic structure information via

\begin{equation}
    \bm{h}^{(\mathcal L+1)}_i = \bm{h}^{(\mathcal L)}_i + \bm{W}_{\mathrm{out}}
    \left[
        \sum_{j\in\mathcal{N}_i}
        f_{\mathrm{node}}^{\mathcal L}
        \left\{
            \left( \bm{W}_{i} \bm{h}^{(\mathcal L)}_i + \bm{b}_{i} \right) \parallel \left( \bm{W}_{j} \bm{h}^{(\mathcal L)}_j + \bm{b}_{j} \right)
        \right\}
        \circ 
        \bm{W}_{\Psi} \Psi_{ji}^{(\mathcal{L})}
    \right].
\end{equation}

\subsection{Output block}
The node embedding vector $\bm{h}_i^{(L)}$, updated through the $L$-layer interaction blocks, is finally passed to the output block.
In the output block, the node embedding is transformed by a neural network to obtain a atomic scalar value $t_i$.
These outputs are summed or averaged across the graph to obtain the final predicted scalar value $t$, depending on whether the physical property value is extensive or intensive.

\section{Experiments and results}

\begin{table}[tb]
  \caption{MAE on Perovskite for the test set. The results are averaged over five randomly created splits.
  SchNet was trained using SchNetPack \cite{schnetpack}, and M3GNet was also trained using publicly available code \cite{m3gnet-github}.
  All models were trained from scratch.
  Ablation studies were also performed on LCAONet to verify the use of spherical Bessel functions developed on DimeNet as radial basis functions, the use of the same basis functions for all elemental species without optimizing the atomic basis functions, and the use of no electron configuration information for node embedding.
  When the interaction layers are aligned to 4 layers and the graph embedding dimension to 128 dimensions, the number of parameters is 1,147,174 for M3GNet, while 869,185 for LCAONet, achieving the better accuracy with about 75\% of the number of parameters.
  }
  \label{table:pero}
  \centering
  \begin{tabular}{lc}
    \toprule
    model & MAE (meV/atom) \\
    \midrule
    OGCNN result from \cite{OGCNN2020-og} & 50 \\
    SchNet & 51.8 $\pm$ 1.2 \\ 
    M3GNet & 38.5 $\pm$ 0.5 \\ 
    \textbf{LCAONet} & \textbf{37.5} $\pm$ 1.3 \\  
    LCAONet with spherical Bessel function as $\bm{e}_{RBF}^{(ij)}$  & 41.3  $\pm$ 1.3 \\  
    LCAONet No optimized basis & 37.9  $\pm$ 1.6 \\  
    LCAONet No electronic info into $\bm{h}^{(0)}$ & 38.9 $\pm$ 2.0 \\  
    \bottomrule
  \end{tabular}
\end{table}


The accuracy of the model was verified for the periodic systems with a various kind of elemental species.
The model hyperparameters and training settings see the supplemental material.

\paragraph{Models.}
For comparison, we used the state-of-the-art model for the crystalline dataset such as SchNet\cite{schnet-2018}, M3GNet\cite{M3GNet_2022-up} and OGCNN\cite{OGCNN2020-og}.
M3GNet has been developed as a universal model that can handle a wide variety of crystalline materials with various elemental species on the periodic table.
Comparisons were also made with OGCNN, which combines the orbital representation with MPNN.

\paragraph{Cubic-Perovskite - periodic dataset.}
Cubic-Perovskite (space group $Pm\bar{3}m$) dataset\cite{perovskite} is created to search for efficient photoelectrochemical cells (PEC), consisting of $18,928$ different structures and formation energy by calculated with DFT, including hypothetical materials. This dataset holds 56 elemental species, including heavy elements with atomic numbers up to 83, and LCAONet's performance was tested in terms of its ability to handle a wide variety of elemental species for the periodic system.
Training was split into $80\%, 10\%, 10\%$ each for training, validation, and test set, following prior research \cite{OGCNN2020-og}, and SchNet, M3GNet and LCAONet were learned from scratch with using open source codes \cite{schnetpack, m3gnet-github}.
Table \ref{table:pero} reports the mean absolute error (MAE) for each model, with LCAONet achieving state-of-the-art accuracy.
Accurate prediction of heavy elements and hypothetical structures is an extremely important property for models to have in the search for new materials, and it is clear that LCAONet is capable of handling a large number of elemental species.
Note that M3GNet is being developed as a model covering the periodic table.
In addition, While the OGCNN achieved improved accuracy by explicitly incorporating the valence orbital interaction into the MPNN, it is believed that the model in this study was able to incorporate orbital information more efficiently through message passing.
When the hyperparameter is aligned with 4 interaction layers and 128 embedded dimensions of the graph, the number of parameters is 1,147,174  for M3GNet while 869,185 for LCAONet, which is better accuracy with about 75\% of the number of parameters, indicating that the developed basis functions and message-passing are superior in parameter efficiency.
Note that M3GNet uses spherical Bessel functions as the radial basis functions.


\paragraph{Ablation study.}
Individual ablation studies were performed on the Perovskite dataset to verify whether the developed basis functions, the optimization of basis functions for each elemental species, and the embedding of electronic structure information contributed to the improved accuracy.
Using the spherical Bessel basis functions proposed in DimeNet as the radial basis function $\bm{e}_{RBF}^{(ij)}$, the error increased by 10\% over the original LCAONet, suggesting that the basis functions developed in this study are effective in improving accuracy.
The accuracy also deteriorated when the radial basis functions were not optimized and the same functional form was used for all elemental species.
Note that optimizing the basis functions according to elemental species limits the geometric representation.
The fact that optimization of basis functions leads to improved accuracy despite these limitations means that it is important to properly represent the geometric spread for each elemental species.
The LCAONet error increased by about 2\% when electronic structure information was not used in the generation of the initial node embedding vectors.
These results indicate that the optimization of basis functions and embedding of electronic structure information can provide an effective inductive bias to the model.

\section{Conclusion}
In this work, we developed physically motivated atomic radial basis functions and introduced electronic structure message-passing, which encompasses orbital information, in order to make accurate predictions for a wide variety of elemental species on the periodic table.
The geometrical features of the graph are vectorized by the radial basis functions of hydrogen-like atoms for the inter-atomic distances and by the spherical harmonic basis for the directional information. It was shown that by optimizing the form and number of basis functions according to the electron configuration of the ground state of the atoms, it was possible to propagate a message that expresses an appropriate spatial extent depending on the elemental species.
Furthermore, we showed that message-passing can include physicochemically rich representations by initializing the embedding vectors of nodes and coefficients using electronic structure information.
We constructed LCAONnet, an MPNN that encapsulates these improved methods, and achieved state-of-the-art accuracy for the Cubic perovskite dataset, which contain various elemental species.
Predictive models that can handle arbitrary elemental species are essential in the search for new materials.
In the future, more detailed orbital information, such as spin degeneracy and crystal field splitting, can be included in the message-passing to cover a wider variety of materials.
In addition, our model uses only $E(3)$ invariant geometrical features, and we believe that extending the equivariant MPNN to include such physicochemical information would also be useful.

\begin{ack}
This research was supported by MEXT (Nos 19H00818, and 19H05787) and JST-CREST (JPMJCR1993).
The authors of this work take full responsibilities for its content.
\end{ack}

\bibliographystyle{unsrt}

\small{
    \bibliography{main}
}

\end{document}